
\input phyzzx
\input tables

\sequentialequations

\overfullrule=0pt
\catcode`\@=11
\def\mm{matrix model}

\def \F{\phi}

\def \P{\psi}
\def \D{{\delta}}

\def\NP{{\it Nucl. Phys.\ }}

\def\PL{{\it Phys. Lett.\ }}
\def\PR{{\it Phys. Rev.\ }}

\def\IJMP{{\it Int. Jour. Mod. Phys.\ }}

\def\PRep{{\it Phys. Rep.\ }}

\def\lc{light-cone}
\def\lcq{light-cone quantization}

\def\eqaligntwo#1{\null\,\vcenter{\openup\jot\m@th
\ialign{\strut\hfil
$\displaystyle{##}$&$\displaystyle{{}##}$&$\displaystyle{{}##}$\hfil
\crcr#1\crcr}}\,}
\catcode`\@=12

\REF\Wilson{K. G. Wilson, \PR {\bf D10} (1974) 2445.}
\REF\Stan{For a review, see S. J. Brodsky and H.-C. Pauli,
``Light-Cone Quantization of Quantum Chromodynamics'',
Lectures at the 30th Schladming Winter School in
Particle Physics, SLAC-PUB-5558 (1991).}
\REF\Brod{H.-C. Pauli and S. Brodsky, \PR {\bf D32} (1985) 1993, 2001;
K. Hornbostel, S. Brodsky and H.-C. Pauli, \PR {\bf D41} (1990) 3814;
for a good review, see K. Hornbostel, Ph. D. thesis, SLAC report
No. 333 (1988).}
\REF\Tang{A. C. Tang, S. J. Brodsky, and H. C. Pauli,
\PR {\bf D44} (1991) 1842; D. Mustaki, S. Pinsky, J. Shigemitsu,
and K. Wilson, \PR {\bf D43} (1991) 3411. }
\REF\DK{S. Dalley and I. R. Klebanov, \PR {\bf D47} (1993) 2517.}
\REF\GH{G. `t Hooft, \NP {\bf B72} (1974) 461.}
\REF\qcd{G. 't Hooft, \NP {\bf B75} (1974) 461.}
\REF\KS{I. R. Klebanov and L. Susskind, \NP {\bf B309} 175 (1988).}
\REF\mm{For a review, see A. A. Migdal, \PRep {\bf 102} (1983) 199.}
\REF\Thorn{C. Thorn, \PL {\bf 70B} (1977) 85; \PR {\bf D17} (1978)
1073.}
\REF\dkut{D. Kutasov, ``Two Dimensional QCD coupled to Adjoint
Matter and String Theory'', EFI-93-30, {\tt hep-th/9306013}.}
\REF\bg{I. Bars and M. Green, \PR {\bf D17} (1978) 537.}
\REF\KKS{M. Karliner, I. Klebanov and L. Susskind,
\IJMP {\bf A3} (1988) 1981.}
\REF\ext{R. Bulirsch and J. Stoer, {\it Numer. Math.} {\bf 6}
(1964) 413; M. Henkel and G. Sch\"utz, {\it J. Math. Phys. A: Math.
Gen.} {\bf 21} (1988) 2617.}
\REF\I{I. Klebanov, Lecture at the YITP Workshop on Quantum Gravity,
Uji, Japan (1992). }
\REF\Polyakov{A. M. Polyakov, ``Two-dimensional quantum
gravity. Superconductivity  at high $T_c$'', in
Fields, Strings and Critical Phenomena, North Holland,
Amsterdam, 1989.}
\REF\Zh{A. Zhitnitsky, \PL {\bf 165B} (1985) 405.}
\REF\Wein{F. Butler, H. Chen, J. Sexton, A. Vaccarino and D. Weingarten,
``Hadron Mass Predictions of the Valence Approximation to Lattice QCD'',
IBM-RC-18617, {\tt hep-lat/9212031}.}
\REF\DG{D. J. Gross, ``Two Dimensional QCD as a String Theory'', PUPT-1356,
{\tt hep-th/9212149};
D. J. Gross and W. Taylor, ``Two Dimensional QCD is a String Theory'',
PUPT-1376, LBL-33458, UCB-PTH-93/02, {\tt hep-th/9301068};
Twists and Wilson Loops in the String Theory of Two Dimensional QCD'',
CERN-TH.6827/93, PUPT-1382, LBL-33767, UCB-PTH-93/09,
{\tt hep-th/9303046}.}
%
%
%
%
%

\font\cmss=cmss10 \font\cmsss=cmss10 at 7pt
\def\IZ{\relax\ifmmode\mathchoice
   {\hbox{\cmss Z\kern-.4em Z}}{\hbox{\cmss Z\kern-.4em Z}}
   {\lower.9pt\hbox{\cmsss Z\kern-.4em Z}}
   {\lower1.2pt\hbox{\cmsss Z\kern-.4em Z}}\else{\cmss Z\kern-.4em Z}\fi}

\def\eqaligntwo#1{\null\,\vcenter{\openup\jot\m@th
\ialign{\strut\hfil
$\displaystyle{##}$&$\displaystyle{{}##}$&$\displaystyle{{}##}$\hfil
\crcr#1\crcr}}\,}
\catcode`\@=12

\def\oh{{1 \over 2}}
\def\b{\beta}
\def\a{\alpha}

\def\half{{1\over 2}}
\def\d{\dagger}

\def\pa{\partial}

\def\dj{\hbox{d\kern-0.347em \vrule width 0.3em height 1.252ex depth
-1.21ex \kern 0.051em}}

\nopagenumbers
\vsize=8.9in
\hsize=6.5in

{\baselineskip=16pt
\line{\hfil PUPT-1413}
\line{\hfil IASSNS-HEP-93/42}
\line{\hfil {\tt hep-th/9307111}}
}

\bigskip

\title{{\bf 1+1-Dimensional Large $N$
QCD coupled to Adjoint Fermions}}

\bigskip

\centerline{{\caps Gyan Bhanot}}
\medskip
\centerline{
\vbox{\hsize2.7in
\centerline{\sl School of Natural Sciences}
\centerline{\sl Institute for Advanced Study}
\centerline{\sl Princeton, New Jersey 08540} }
\vbox{\hsize0.4in
\centerline{}
\centerline{and}
\centerline{}}
\vbox{\hsize2.7in
\centerline{\sl Thinking Machines Corporation}
\centerline{\sl 245 First Street}
\centerline{\sl Cambridge, MA 02142}}
}
\bigskip
\centerline {{\caps Kre\v simir Demeterfi}\foot{{\rm On leave of absence
from the Ru\dj er Bo\v skovi\'c Institute, Zagreb, Croatia}}
and {\caps Igor R. Klebanov} }
\centerline{\sl Joseph Henry Laboratories}
\centerline{\sl Princeton University}
\centerline{\sl Princeton, New Jersey 08540}

\bigskip
\abstract
We consider 1+1-dimensional QCD coupled to Majorana fermions in the
adjoint representation of the gauge group $SU(N)$. Pair creation of
partons (fermion quanta) is not suppressed in the large-$N$ limit,
where the glueball-like bound states become free. In this limit the
spectrum is given by a linear \lc\ Schr\" odinger equation, which we
study numerically using the discretized \lcq. We find a discrete
spectrum of bound states, with the logarithm of the level
density growing approximately linearly with the mass.
The wave function of a typical excited
state is a complicated mixture of components with different parton numbers.
A few low-lying states, however, are surprisingly close to being
eigenstates of the parton number, and their masses can be
accurately calculated by truncated diagonalizations.
\vfill
\line{7/93\hfill}
\endpage

\pagenumbers
\centerline{\caps 1. Introduction}
\bigskip

QCD has become universally accepted as the correct
theory of strong interactions,
on the basis of a large body of experimental and theoretical evidence.
However, there are few reliable non-perturbative calculations that can
be carried out starting from first principles. One of the most explored
approaches to non-perturbative QCD has been the lattice gauge theory
[\Wilson].
In fact, numerical studies of pure glue theory appear to be close to
the continuum limit, and significant progress is being made on various
versions of lattice QCD with quarks. It is important, however, to look
for other non-perturbative methods, in the hope that they will lead to
new qualitative and quantitative insights. One such approach makes use
of the \lc\ quantization and subsequent numerical diagonalization of
the \lc\ Hamiltonian [\Stan]. It may provide tools
for calculating the hadron spectrum,
as well as the wave functions in the infinite momentum frame, the decay
amplitudes and the interaction cross-sections. Among its other advantages
is the ability to introduce chiral fermions without obvious
complications. This approach has been successfully applied to
QCD and other model field theories in 1+1 dimensions [\Brod], and is
currently being generalized to 3+1 dimensional theories [\Tang].
In this paper we consider an application of the \lc\ approach
to a new type of models [\DK] where, we
feel, it is particularly well suited.

As proposed by `t Hooft, QCD simplifies when generalized to a large
number of colors $N$ [\GH]. When combined with the \lc\ quantization,
this simplification becomes particularly striking: in the $N\to\infty$
limit meson and glueball wave functions are solutions of {\it linear}
\lc\ Schr\"odinger equations [\qcd, \KS]. This is related to the fact that
mesons and glueballs become free in the large-$N$ limit. The linearity
of the equations, however, is a special property of the \lcq. Recall,
for comparison, that the loop equations remain non-linear
in the large-$N$ limit [\mm].

In this paper we will study the spectrum of such linear \lc\
Schr\"odinger equation for a particular model, 1+1 dimensional
large-$N$ QCD coupled to matter in the adjoint representation of
$SU(N)$ [\DK]. This model is far more complex than the large-$N$ QCD
coupled to quarks in the fundamental representation,
where `t Hooft derived and numerically solved the bound state
equation for mesons [\qcd]. The quanta of the adjoint matter resemble gluons
in that there are two color flux tubes attached to each quantum.
The resulting glueball-like
bound states may contain any number of quanta connected into
a closed string by the color flux tubes (see Fig.~1). As we will see, the
eigenstates are generally complex mixtures of such strings with
different numbers of partons. This should be contrasted with
the `t Hooft model where, due to the absence of transverse gluons,
all the meson bound states have the structure of a quark and an
antiquark connected by a color flux tube. Our introduction of
adjoint matter has the purpose of imitating some transverse gluon
effects. In fact, if we
dimensionally reduce 2+1-dimensional gauge theory,  the zero
mode of the transverse gluon field acts as the adjoint matter field
coupled to 1+1-dimensional QCD. Therefore, this model seems to be
the simplest setting where one can study some genuine QCD effects,
such as the pair creation of partons. We will argue that the
\lcq\ supplemented with a regulator in the form of discretized
longitudinal momenta [\Brod, \Thorn]
allows one to extract significant amount
of physical information about the large-$N$ theory.

Consider the pure glue $SU(N)$ gauge theory in 2+1 dimensions,
$$ S=-{1\over 4g_3^2}\int d^3 x \Tr F_{\mu\nu}F^{\mu\nu}
\ . \eqn\eq$$
If one of the spatial dimensions is made compact,
$y\sim y + L$, then as $L\to 0$ we may ignore the dependence of
fields on $y$, \ie\ $\partial A^\mu/\partial y =0$.
The action then reduces to
$$ S_{\rm sc} = \int d x^0 dx^1
 \Tr \left[ \oh D_{\a}\F D^{\a}\F
-{1\over 4 g^{2}} F_{\a \b}F^{\a \b}\right], \eqn\action$$
where $g^2=g_3^2/L$, and $\phi (x^0, x^1)=A_y/g$
is a traceless $N\times N$ Hermitian matrix field, whose
covariant derivative is given by $D_{\a}\phi = \pa_{\a}\phi
+i[A_{\a}, \phi]$. Therefore, $\phi$ represents the remnants of the
transverse gluon degrees of freedom.
If we choose the \lc\ gauge $A_-=0$ and add a mass term for $\phi$,
we obtain
$$S_{\rm sc}=\int dx^{+}dx^{-} \Tr \left[\pa_{+}\F\pa_{-}\F -\half m^2\F^2
+{1\over 2 g^2} (\pa_{-}A_{+})^2 + A_{+}J^{+} \right]\,\,, \eqn\mact$$
where $x^{\pm}=(x^0\pm x^1)/\sqrt 2$,
and the longitudinal momentum current $J^{+}_{ij} = i[\F,\pa_{-}\F]_{ij} $.
The mass term for $\phi$, which does not destroy the 1+1 dimensional
gauge invariance, is necessary to absorb the logarithmically divergent
mass renormalization. The \lcq\ and the spectrum of the theory
\mact\ were considered in Ref.~[\DK]. However, in the numerical
diagonalization no proper account was taken of the divergent mass
renormalization. \foot{We thank D. Kutasov for helpful discussions
on this issue.}
The bare mass was held fixed, and hence all
the bound state masses were diverging in the continuum limit.
A proper numerical diagonalization, where the renormalized mass is held
fixed, is in progress, and we hope to report on it in the future.

In the present paper we will examine, instead, a simpler model
where the adjoint scalar is replaced by an adjoint Majorana fermion
[\DK].  Thus, we obtain a 1+1-dimensional gauge theory coupled to the
zero mode of a transverse gluino. The bound states again are built
of any number of partons connected into a closed string by color flux
tubes. The new feature is that the bound states are fermions or bosons
depending on whether the number of partons is even or odd.
This theory has the advantage of
being perfectly finite; moreover, it is supersymmetric for a special
value of the fermion mass [\dkut]. In Ref.~[\DK], S. Dalley and one of the
authors carried out the \lcq\ of this theory, and began a numerical
investigation of the low-lying spectrum. Here we continue this program
with further analytical and numerical results.

\bigskip
\centerline{\caps 2. Light-cone quantization}
\bigskip

Consider $N^2-1$ Majorana (real) fermions which transform in the adjoint
representation of $SU(N)$. They can be combined into a traceless
Hermitian matrix $\Psi_{ij}$. Upon gauging the $SU(N)$ symmetry
we obtain the action
$$ S_{\rm f} = \int d^2 x \Tr \left[
i\Psi^T\gamma^{0} \gamma^{\a} D_{\a} \Psi -m\Psi^{T}\gamma^{0} \Psi
-{1\over 4 g^{2}} F_{\a \b}F^{\a \b}\right]\,\,, \eqn\faction$$
where the transposition acts only on the Dirac indices,
and the covariant derivative is defined by
$D_{\a}\Psi = \pa_{\a}\Psi
+i[A_{\a}, \Psi]$. The fermion field
$\Psi_{ij}=2^{-1/4} {\psi_{ij}\choose \chi_{ij}}$
is a two-component spinor, where
$\chi$ and $\P$ are  traceless Hermitian
$N\times N$ matrices of Grassmann variables.
Choosing the light-cone gauge $A_-=0$, and
the representation $\gamma^0=\sigma_2$, $\gamma^1=i\sigma_1$,
we find the action
$$S_{\rm f}=\int dx^{+}dx^{-} \Tr \left[
i\psi \pa_{+} \psi + i\chi \pa_{-} \chi -i\sqrt 2 m \chi \psi
+{1\over 2 g^2} (\pa_{-}A_{+})^2 + A_{+}J^{+} \right]\,\,,\eqn\mact$$
where the longitudinal momentum current is now of the form
$J^{+}_{ij} = 2\psi_{ik} \psi_{kj}$.

In the \lcq\ $x^+$ is treated as
the time, and the canonical anti-commutation relations are imposed
at equal $x^+$,
$$\{\psi_{ij}(x^{-}), \psi_{kl}(y^{-})\} = \half\,
\delta(x^{-}-y^{-})\bigl ( \delta_{il} \delta_{jk}-
{1\over N}\delta_{ij} \delta_{kl}\bigr)\,\,.\eqn\anticomm$$
The action does not contain time derivatives of
$A_+$ and $\chi$, and these non-dynamical
fields can be eliminated by their constraint equations.
As a result, the light-cone components of total momentum can be
expressed in terms of $\psi$ only,
$$ \eqalign{&P^{+} =  \int dx^{-} \Tr\,
[i\psi \pa_{-} \psi]\ ,\cr
&P^{-} = \int dx^{-} \Tr \left[-{im^2 \over 2}
\psi {1\over \pa_{-}} \psi -  \half g^2  J^{+} {1\over \pa_{-}^{2}} J^{+}
\right]\ .\cr }\eqn\fminus$$
Our goal is to solve the eigenvalue
problem
$$ 2P^+ P^- |\Phi\rangle= M^2 |\Phi\rangle\,. \eqn\ev$$
Since $[P^+, P^-]=0$, $|\Phi\rangle$ is a simultaneous eigenstate
of $P^+$ and $P^-$. In practice it is easy to ensure that $|\Phi\rangle$
carries a definite $P^+$, but the subsequent solution of Eq.~\ev\ is
highly non-trivial.
All the physical states must also satisfy the zero-charge constraint,
$$\int dx^- J^+  |\Phi\rangle = 0\ ,
\eqn\zc$$
arising from integration over the zero-mode of $A_+$, which acts as
a Lagrange multiplier.

In order to make Eq.~\ev\ explicit, we introduce
the mode expansion
$$\psi_{ij}(x^-) = {1\over 2\sqrt\pi} \int_{0}^{\infty} dk^{+}
\left(b_{ij}(k^{+}){\rm e}^{-ik^{+}x^{-}} +
b_{ji}^{\d}(k^{+}){\rm e}^{ik^{+}x^{-}}\right )\ .\eqn\mode $$
{}From Eq.~\anticomm\
it follows that
$$\{b_{ij}(k^{+}), b_{lk}^{\d}(\tilde{k}^{+})\} =
\delta(k^{+} - \tilde{k}^{+})
(\delta_{il} \delta_{jk}-{1\over N}\delta_{ij} \delta_{kl})\,\,.
\eqn\modac$$
In terms of the oscillators, Eq.~\fminus\ assumes the form
$$P^+ = \int_{0}^{\infty} dk\ k\, b_{ij}^{\d}(k)b_{ij}(k)\ ,\eqn\fp$$
$$\eqalign{&P^{-} = {m^2\over 2}\, \int_{0}^{\infty}
{dk\over k} b_{ij}^{\d}(k)
b_{ij}(k) +{g^2 N\over \pi} \int_{0}^{\infty} {dk\over k}\
C(k) b_{ij}^{\d}(k)b_{ij}(k) \cr
&+ {g^2\over 2\pi} \int_{0}^{\infty} dk_{1} dk_{2} dk_{3} dk_{4}
\biggl\{ A(k_i) \D (k_{1}+k_{2}-k_{3}-k_{4})
b_{kj}^{\d}(k_{3})b_{ji}^{\d}(k_{4})b_{kl}(k_{1})b_{li}(k_{2})  \cr
& + B(k_i) \D(k_{1} + k_{2} +k_{3} -k_{4})
(b_{kj}^{\d}(k_{4})b_{kl}(k_{1})b_{li}(k_{2})
b_{ij}(k_{3})-
b_{kj}^{\d}(k_{1})b_{jl}^{\d}(k_{2})
b_{li}^{\d}(k_{3})b_{ki}(k_{4})) \biggl\} \cr }\eqn\fpminus$$
where
$$\eqalign{&A(k_i)= {1\over (k_{4}-k_{2})^2 } -
{1\over (k_{1}+k_{2})^2}\ , \cr
&B(k_i)= {1\over (k_{2}+k_{3})^2 } - {1\over (k_{1}+k_{2})^2 }\ , \cr
&C(k)= \int_{0}^{k} dp \,\,{k\over (p-k)^2}\ , \cr }\eqn\coeffer$$
and we have dropped the superscripts $+$ on $k_i$ for brevity.
The mass renormalization proportional to $C$ arises from the normal ordering
of the quartic term in $P^-$. If one uses `t Hooft's principal value
prescription, then $C(k)=-1$, so that the mass renormalization is finite.
Other prescriptions, however, such as the one we will use, render
$C(k)$ linearly divergent. It is important to keep in mind,
however, that $C$ is not a physical quantity because it enters in
the mass of a colored object. The physical quantities are the masses
of the colorless bound states, and they must be independent of
which consistent prescription is used to define $P^-$. This was the case for
the `t Hooft model [\bg], and we expect the same to be true here.

An important advantage of the \lcq\ is that the oscillator vacuum
satisfies
$$ P^+ |0\rangle=0;\qquad\qquad P^- |0\rangle=0\ .\eqn\eq$$
Other states in the Fock space are constructed by acting with creation
operators $b^\dagger_{ij}$ on the vacuum. The zero-charge condition
\zc\ requires that all the color indices be contracted.
Therefore, we look for bosonic eigenstates of Eq.~\ev\ in the form
$$\eqalign{|\Phi_{\rm b}(P^+) \rangle
=&\sum_{j=1}^\infty \int_0^{P^+} dk_1 \ldots dk_{2j} \,
\delta\Bigl(\sum_{i=1}^{2j} k_i-P^+\Bigr) \cr &
f_{2j} (k_1, k_2, \ldots, k_{2j})
N^{-j} \Tr \,[b^{\d}(k_1)\ldots b^{\d}(k_{2j})] |0 \rangle\ . \cr }
\eqn\bw$$
This state is trivially an eigenstate of $P^+$, and the problem is
to ensure that it is an eigenstate of $P^-$. Similarly, the fermionic
states are of the form
$$\eqalign{|\Phi_{\rm f}(P^+) \rangle
=&\sum_{j=1}^\infty \int_0^{P^+} dk_1 \ldots dk_{2j+1} \,
\delta\Bigl(\sum_{i=1}^{2j+1} k_i-P^+\Bigr) \cr &
f_{2j+1} (k_1, k_2, \ldots, k_{2j+1})
N^{-j-1/2} \Tr\,[b^{\d}(k_1)\ldots b^{\d}(k_{2j+1})] |0 \rangle\ . \cr }
\eqn\fw$$
Due to the fermionic statistics of the oscillators, the wave functions
have cyclic symmetry
$$f_i (k_2, k_3, \ldots, k_i, k_1)=(-1)^{i-1} f_i (k_1, k_2,\ldots, k_i)\,\,.
\eqn\cyclic$$

The increased complexity of
the coupling to adjoint matter arises mainly from the fact that the
eigenstates are mixtures of states with different numbers of
partons. This can be traced to the presence of pair
production and pair annihilation terms in $P^-$. One easily checks
that these appear in the leading order of the $1/N$ expansion,
provided that $g^2 N$ is kept fixed in the large-$N$ limit.
Indeed, for the model with adjoint matter an extra pair of partons can be
produced {\it inside} a color singlet. Furthermore, the terms in
$P^-$ that take one color singlet into two are suppressed by
$1/N$. Therefore, our bound states are stable in the large-$N$
limit, and their wave functions satisfy {\it linear} eigenvalue
equations [\KS, \DK]. These equations are not hard to write down explicitly.
Upon introducing longitudinal momentum fractions $x_i=k_i^+/P^+$,
we find the following set
of coupled integral equations by acting on states \bw\ and \fw\ with $P^-$,
$$\eqalign{& M^2 f_i (x_1, x_2, \ldots, x_i)=
{m^2\over x_1}
f_i (x_1, x_2, \ldots, x_i) +{g^2 N\over\pi (x_1+x_2)^2}
\int_0^{x_1+x_2} dy f_i (y, x_1+ x_2-y, x_3, \ldots, x_i)
\cr
&+{g^2 N\over\pi}\int_0^{x_1+x_2} {dy \over (x_1-y)^2}
\bigl [f_i (x_1, x_2, x_3, \ldots, x_i)-
f_i (y, x_1+ x_2-y, x_3, \ldots, x_i) \bigr ]\cr
&+{g^2 N\over\pi}\int_0^{x_1} dy \int_0^{x_1-y} dz
f_{i+2} (y,z, x_1-y-z, x_2, \ldots, x_i)
\left [{1\over (y+z)^2}-{1\over (x_1-y)^2}\right ]\cr
&+{g^2 N\over\pi}
f_{i-2} (x_1+ x_2+x_3, x_4, \ldots, x_i)
\left [{1\over (x_1+x_2)^2}-{1\over (x_2+x_3)^2}\right ]\cr
\noalign{\vskip 0.2cm}
& \pm {\rm cyclic~ permutations~ of~} (x_1, x_2, \ldots, x_i)\,\,. \cr
}\eqn\LS$$
For odd $i$ all cyclic permutations enter with positive sign,
while for even $i$ they enter with alternating signs. This is related to
the cyclic symmetry \cyclic. In the second line of Eq.~\LS\ the Coulomb
double pole is partly compensated by the zero of the numerator
at $y=x_1$. Therefore, the integral is finite in the principal value
sense, and the equation contains no ambiguity. The same holds true for
the `t Hooft equation (33).

Eq. \LS\ possesses a $\IZ_2$ symmetry $T$ [\dkut]. For any fermionic (odd $i$)
eigenstate,
$$f_i (x_1, x_2, \ldots, x_i)=T (-1)^{(i-1)/2} f_i (x_i, \ldots, x_2, x_1)
\ ,\eqn\eq $$
while for any bosonic (even $i$) eigenstate,
$$f_i (x_1, x_2, \ldots, x_i)=T (-1)^{i/2} f_i (x_i, \ldots, x_2, x_1)
\ . \eqn\eq$$
The $\IZ_2$ quantum number
$T$ has two possible values, 1 and $-1$. In terms of the original
field, $T: \psi_{ij} \to \psi_{ji}$, which obviously leaves $P^\pm$
invariant [\dkut]. Physically, every bound state can be thought of as a
superposition of
oriented closed strings, and the quantum number $T$ describes the
transformation property under a reversal of orientation.
\endpage

\bigskip
\centerline{\caps 3. The discretized approximation}
\bigskip

The system of equations \LS\ involves an infinite number of
multivariable functions. The complexity of the adjoint matter
model is evidently much greater than that of the `t Hooft model,
where each bound state is specified by a single function of one
variable. Even there, however, one needs to resort to numerical methods
to find the eigenvalues of the linear \lc\ Schr\"odinger equation.
Here we follow a similar strategy and replace the continuum
equations \LS\ by a sequence of discretized approximations, such that
the eigenvalues of the discretized problems eventually converge to the
eigenvalues of \LS.
In the \lc\ quantization, a simple discretized approximation is
obtained by replacing
the continuous momentum fractions $x$ by a discrete set $n/K$,
where $n$ are odd positive integers, and
the positive integer $K$ is sent to infinity as the cut-off
is removed [\Thorn,\Brod].
Thus, the functions $f_i (x_1, x_2, \ldots, x_i)$ are
replaced by finite collections of numbers which specify their values
at the discrete set of $x$, and
$$\int_0^1 dx \to {2\over K}\sum_{{\rm odd~}n>0}^K\,\,.
$$
Moreover, the constraint $\sum_{j=1}^i x_j=1$ eliminates all states
with over $K$ partons, so that the discretized eigenvalue problem
becomes finite-dimensional. Any given eigenstate
of the continuous problem should be well approximated by the
discretizations with large enough $K$, although in practice the
convergence may be very slow for highly excited states.

An equivalent way to describe our cut-off is in
terms of the discretized \lcq\ [\Brod]. There one makes $x^-$ compact and
imposes anti-periodic boundary conditions,
$\psi_{ij}(x^-)=-\psi_{ij}(x^-+2\pi L)$.
\foot{In ref. [\DK] periodic boundary conditions were used
instead. Although for either choice of the boundary conditions
the theory eventually converges to the limit of continuous $k^+$, we find that
the convergence is appreciably faster for the
anti-periodic boundary conditions.} Therefore, $k^+$ is restricted
to discrete values $n/(2L)$ where $n$ are odd positive integers.
The total light-cone momentum is $P^+=K/(2L)$, where $K$ is odd for
the fermionic bound states whose wave functions
are anti-periodic in $x^-$, and $K$ is even for
the bosonic bound states which are periodic in $x^-$.
The mode expansion can now be written as
$$\psi_{ij}(x^-) ={1 \over \sqrt{4\pi}} \sum_{{\rm odd~}n>0}
\left (B_{ij}(n){\rm e}^{-iP^{+}nx^-/K} + B_{ji}^{\d}(n)
{\rm e}^{iP^{+}nx^-/K} \right )\ ,
\eqn\dm$$
with the oscillator algebra
$$
\{ B_{ij}(n),B_{lk}^{\d}(n')\} =
\delta_{n n'}
\bigl (\delta_{il}\delta_{jk}-
{1\over N}\delta_{ij} \delta_{kl}\bigr)\ .\eqn\modecomm$$
The matrix that has to be diagonalized can be constructed in
terms of the oscillators,
$$2P^+P^{-} = {g^2 N\over \pi}~K\left(x V +T\right)
\ ,\eqn\fham$$
where $x={\pi m^2\over g^2 N}$ is
the dimensionless parameter. The mass term is
$$V=\sum_n {1\over n}\,
B^{\d}_{ij}(n)B_{ij}(n)\,\,, \eqn\pot$$
while the term generated
by the gauge interaction is
$$\eqalign{&T= 4\sum_n
B^{\d}_{ij}(n)B_{ij}(n) \sum_{m}^{n-2} {1\over (n-m)^2 } +\cr
& {2\over N}\sum_{n_i} \biggl\{
\delta_{n_1+n_2, n_3+n_4}
\left[ {1\over (n_{4}-n_{2})^2 }
-{1\over (n_{1} + n_{2})^2 } \right]
B_{kj}^{\d}(n_{3})B_{ji}^{\d}(n_{4})B_{kl}(n_{1})B_{li}(n_{2})  \cr
& + \delta_{n_1+n_2+n_3,n_4}
\left[{1\over (n_{3}+n_{2})^2 } -
{1\over (n_{1}+n_{2})^2 } \right]\cr
&\bigl(
B_{kj}^{\d}(n_{4})B_{kl}(n_{1})B_{li}(n_{2}) B_{ij}(n_{3})
-B_{kj}^{\d}(n_{1})B_{jl}^{\d}(n_{2}) B_{li}^{\d}(n_{3})B_{ki}(n_{4})
\bigr) \biggr\}\,\,. \cr }\eqn\fmess$$
All the summations above are restricted to positive odd integers.

In order to perform the diagonalization, we may consider a basis of
states normalized to 1 in the large $N$ limit,
$${1\over N^{i/2}\sqrt{s}}
\Tr [B^{\d}(n_1)\cdots B^{\d}(n_i)] |0\rangle\ ,\qquad\qquad
\sum_{j=1}^i n_j=K \ . \eqn\fstring$$
The states are defined by
ordered partitions of
$K$ into $i$ positive odd integers, modulo cyclic permutations.
If $(n_1,~ n_2,~ \ldots,~ n_i)$ is taken into itself
by $s$ out of $i$ possible cyclic permutations,
then the corresponding state receives a
normalization factor $1/\sqrt{s}$. In the absence of special
symmetries, $s=1$.
For even $i$, however,
some such states vanish due to the fermionic statistics of the oscillators:
all partitions of $K$ where $i/s$ is odd do not
give rise to states.

In actual calculations it is advantageous to consider separately
the even and odd sectors under
$T: B^\dagger_{ij}(n) \to B^\dagger_{ji}(n)$. The states that
carry a definite quantum number $T$ are in general linear combinations
of the states $\fstring$. Construction and proper normalization of such
states is a combinatorial problem that is easily solved with a computer
program. We will show the solution for a simple example,
setting $K=10$.

In the $T=1$ sector the normalized states are
$$ \eqalign{
&|1\rangle= {1\over N^3}\Tr [B^\dagger(5) B^\dagger (1)
B^\dagger(1)B^\dagger(1)B^\dagger(1)B^\dagger(1)] |0\rangle\ , \cr
&|2\rangle= {1\over N^3}\Tr [B^\dagger(3) B^\dagger (1)
B^\dagger(3)B^\dagger(1)B^\dagger(1)B^\dagger(1)] |0\rangle\ , \cr
&|3\rangle={1\over N^2 \sqrt 2}\bigl (\Tr [B^\dagger (5)B^\dagger (3)
 B^\dagger(1)B^\dagger(1)] +
\Tr [B^\dagger (1)B^\dagger (1) B^\dagger(3)B^\dagger(5)] \bigr )
|0\rangle\ ,\cr
&|4\rangle=
{1\over N}\Tr [B^\dagger(9) B^\dagger (1) |0\rangle\ ,\qquad\qquad
|5\rangle={1\over N}\Tr [B^\dagger (7)B^\dagger (3) ] |0\rangle\ .\cr}
\eqn\eq$$
Here the matrix to be diagonalized is
$$ xV+T=\left (\matrix
{ {26x\over 5} + {7\over 2} & 0 & 0 & 0 & 0 \cr
0 & {14x\over 3}+ {7\over 2} & 0 & 0 & 0 \cr
0 & 0 &  {38x\over 15}+{359\over 144} & -{17\over 72\sqrt 2} &
{4\sqrt 2\over 9} \cr
0 & 0 & -{17\over 72\sqrt 2} & {10x\over 9}+{107\over 72} & -{8\over 9} \cr
0 & 0 & {4\sqrt 2\over 9} & -{8\over 9} &{10x\over 21} + {47\over 18} }
\right ) \ . \eqn\eq$$

In the $T=-1$ sector the normalized states are
$$ \eqalign{
&|1\rangle= {1\over N^4}\Tr [B^\dagger(3) B^\dagger (1)B^\dagger(1)B^\dagger(1)
B^\dagger(1)B^\dagger(1)B^\dagger(1)B^\dagger(1)] |0\rangle\ , \cr
&|2\rangle= {1\over N^3}\Tr [B^\dagger(3) B^\dagger (3)
B^\dagger(1)B^\dagger(1)B^\dagger(1)B^\dagger(1)] |0\rangle\ , \cr
&|3\rangle={1\over N^2 }
\Tr [B^\dagger (7)B^\dagger (1) B^\dagger(1)B^\dagger(1)] |0\rangle\ ,\cr
&|4\rangle={1\over N^2 \sqrt 2}\bigl(\Tr [B^\dagger (5)B^\dagger (3)
 B^\dagger(1)B^\dagger(1)] -
\Tr [B^\dagger (1)B^\dagger (1) B^\dagger(3)B^\dagger(5)] \bigr)
|0\rangle\ ,\cr
&|5\rangle= {1\over N^2}
\Tr [B^\dagger(5) B^\dagger (1)B^\dagger(3)B^\dagger(1) |0\rangle\ ,
\qquad |6\rangle={1\over N^2}
\Tr [B^\dagger (3)B^\dagger (3)B^\dagger(3)B^\dagger(1) ] |0\rangle\ ,\cr }
\eqn\eq$$
and the matrix to be diagonalized is
$$ xV+T=\left (\matrix
{ {22x\over 3} + 5 & 0 & 0 & 0 & 0 & 0 \cr
0 & {14x\over 3}+ {137\over 36} & -{5\over 36} & {3\over 4\sqrt{2}} &
0 & 0 \cr
0 &  -{5\over 36} & {22x\over 7} +{89\over 36} & -{3\over 4\sqrt{2}} &
0 & 0 \cr
0 & {3\over 4\sqrt{2}} &  -{3\over 4\sqrt{2}} &
{38x\over 15}+{247\over 72} & -{11\over 18\sqrt{2}} & {4\sqrt{2}\over 9} \cr
0 & 0 & 0 &  -{11\over 18\sqrt{2}} & {38x\over 15} + {47\over 18}&
-{8\over 9} \cr
0 & 0 & 0 &  {4\sqrt{2}\over 9} & -{8\over 9} &
2x + {37\over 9} }
\right ) \ . \eqn\eq$$

The calculations above were repeated for higher values of $K$
with the help of a computer program.
The number of states increases rapidly with $K$.
Our biggest diagonalization in the fermionic sector was carried
out for $K=25$, where there are 3312 states in the
$\IZ_2$ odd sector and 3400 states in the $\IZ_2$ even sector.
In the bosonic sector we reached $K=24$ where there are
2197 $\IZ_2$ even states and 2141 $\IZ_2$ odd states.

\bigskip
\centerline{\caps 4. The numerical results }
\bigskip

A good numerical procedure is to calculate the spectrum for a fixed
$x$ and a range of values of $K$, and then to extrapolate the results to
infinite $K$, the continuum limit. We will also assume that some bulk
properties of the spectrum can be estimated from the results at a fixed
large $K$. We will be most interested in two special values of $x$,
$x=0$ which corresponds to the limit of massless quanta, and $x=1$
($m^2=g^2 N/\pi$) where the theory is supersymmetric [\dkut].

In Fig.~2(a) we show the spectrum of fermionic states
for $x=0$ and $K=25$,
with the mass plotted vs. the expectation value of the number of
partons, $n$. It is immediately obvious that the density
of states increases rapidly with the mass, and that
almost all the states lie within a band bounded
by two $\langle n \rangle \sim M$ lines.
Below we will try to quantify these effects.

One interesting feature of our results, already noted in Ref.~[\DK]
for smaller $K$, is that for a few low-lying eigenstates the wave
functions are strongly peaked on states with a definite number of
partons. For example, for $K=25$ the ground state has probability
0.99993 to consist of 3 partons, and the first excited state has
probability 0.99443 to consist of 5 partons. As the excitation number
increases, however, the wave functions typically become quantum
superpositions of states with different parton numbers.
It is physically plausible that a typical excited state contains
some number of virtual pairs, and our data supports this expectation.
In order to quantify this effect, we will call a state pure if it
has probability $> 0.9$ to be in one of the number sectors.
Table I shows the total number of states and the number of pure states
in each mass interval of Fig.~2(a). We also show
the expectation value of the number of partons averaged over all states
in each mass interval.
Evidently, a few low-lying states
are pure, while there are no pure states among the high excitations.

As the mass of the quantum increases we expect the pair creation to
become somewhat suppressed. In order to study this effect, we plot
in Fig.~2(b) the spectrum of fermionic states for $x=1$ and $K=25$, and
in Table II we quantify their purity. We find that, indeed, there are
more pure states for low excitation numbers, but for highly excited
states the pair creation again becomes important.
In Figs.~3(a) and 3(b) we show the results for $K=24$  and for
$x=0$ and $x=1$, respectively, in order to
demonstrate that for
the bosonic bound states the qualitative picture is the same.
Table III shows the numerical data for $x=0$.
In Ref.~[\dkut] the spectrum of highly excited states was found
in the approximation where the number changing processes were ignored,
\ie\ all such states were assumed to be pure.
The observed tendency of the excited states to be quantum
superpositions of many number sectors, as well as the
distributions of states in Figs.~2 and 3, do not seem to support
this approximation.
In principle, it is possible
that we have not reached a high enough value of $K$ for our discretized
approximations to detect these pure highly excited states.
We are inclined to believe, however, that a typical highly excited state
does contain a number of virtual pairs of partons.
More discussion of this issue will follow in section 5.

We note a difference in the distribution of states
for $x=0$ and $x=1$. For $x=0$, Figs.~2(a) and 3(a),
states are distributed within a band almost uniformly.
For $x=1$, however, we see an increase in the number
of fermionic states with an average parton number
near $5,7,9,11,\ldots$, Fig.~2(b), and a similar increase
in the number of bosonic states with an average parton
number near $4,6,8,10,\ldots$, Fig.~3(b).
We believe this effect to be related to the
turning on of the mass, and that it gets stronger
as the mass increases. It would be interesting to
investigate this further.

A striking property of Figs.~2 and 3 is the rapid growth of the
density of states
with increasing mass. In Fig.~4 we plot the logarithm of the number of
states vs. the mass for the data in Table I. For a certain range of
masses the graph is approximately linear. The deviation from linearity
for large enough mass is clearly due to the effects of the cut-off.
Our results indicate that the density of states grows roughly
exponentially with the mass, exhibiting the Hagedorn behavior
$$\rho(m)\sim m^\alpha e^{\beta m}\,\,,
\eqn\Hag$$
as suggested in Ref.~[\dkut].
Thus, although the mass spectrum is discrete, it rapidly becomes
virtually indistinguishable from a continuum. From our data we estimate
that the inverse Hagedorn temperature is
$\beta\approx (0.7 - 0.75) \,\sqrt{\pi/ (g^2 N)}$.

Another physical effect that is pronounced in our results
(Tables I-III) is that
the mass increases roughly linearly with the average number of partons.
In Fig.~5  we plot these results for $x=0$ and $K=24$ (Table III).
We will attempt to give a simple heuristic explanation
of this effect. Suppose that
the light-cone Hamiltonian of a glueball-like state containing
on the average $n$ partons is replaced by that of $n$ non-relativistic
particles connected into a closed string by harmonic springs.
It is not hard to see that the ground state energy of such a system,
to be identified with $M^2$, indeed behaves as $\sim n^2$ for sufficiently
large $n$ [\KKS]. Perhaps such a heuristic picture can indeed
help one in a qualitative description of a typical bound state.

Now we need to address the question of convergence towards the continuum
limit. In Fig.~6(a) we show the fermionic and bosonic ground states
for $x=0$, as well as their extrapolation towards infinite $K$;
in Fig.~6(b) we repeat the plot for $x=1$.
We have used the Bulirsch-Stoer algorithm which has proved to
be particularly efficient for extrapolating short series [\ext].
The supersymmetry
of the spectrum for $x=1$ guarantees that the continuum values
of the fermionic and bosonic ground states are equal, and our
extrapolations indeed agree very well.
The numerical values are shown in Tables IV and V.

Since the low-lying states are very pure, they can be well approximated
by truncating the diagonalization to a single parton number sector.
For instance, for $x=0$ and $K=24$ the ground state has probability
0.97366 to consist of 2 partons. We can, therefore, obtain a good upper
bound on its energy by truncating the eigenvalue equations \LS\ to
the two-parton sector. The resulting eigenvalue problem is
$$M^2 \phi (x)=m^2 \phi(x)
\biggl ({1\over x}+{1\over 1-x}\biggr)
+{2g^2 N\over\pi}\int_0^1 dy {\phi(x)-\phi(y)\over (y-x)^2}\,\,,
\eqn\tH$$
where $\phi(x)=f_2(x, 1-x)$. Eq. \tH\ is the `t Hooft equation
with $g^2\to 2g^2$. The doubling of the strength of the interaction term
is due to the presence of two color flux tubes connecting a pair of
partons (in a meson there is only one). Another important new effect
is that, due to the fermionic statistics, $\phi(x)=-\phi(1-x)$.
This forbids half of the eigenstates of the \brk `t Hooft problem,
including the ground state. In particular, for $m=0$ the
$M=0$ solution $\phi(x)=1$ is excluded.
This provides a heuristic argument for the absence of massless bound
states, even as the parton mass $m$ is taken to zero.
A more precise argument will be given in section 5.
An approximation to the bosonic ground
state of the adjoint fermion model is provided by the lowest
antisymmetric wave function, whose eigenvalue is
$M^2\approx 11.76\, g^2 N/ \pi$.
This upper bound
is quite close to the extrapolated value from Fig.~6(a),
which is $M^2\approx 10.7 \,g^2 N/ \pi$.
The lowest antisymmetric eigenstate
of eq. \tH\ for $m^2=g^2 N/\pi$ ($x=1$) has
$M^2\approx 26.56 \,g^2 N/ \pi$, which is a good
upper bound on the extrapolated value
from Fig.~6(b), $M^2\approx 25.9 \,g^2 N/ \pi$.

Since the bosonic ground state is not perfectly pure, the upper bounds
can be improved by including in the truncated diagonalization all the
2-, 4- and 6-bit states. For $x=0$ and $K=24$ the ground state
has probability 0.99998 to be in this sector.
Extrapolating to infinite $K$, we find 0.99995, and therefore
this truncation is highly reliable.
In Table IV we compare the full and truncated calculations.
We have performed the truncated diagonalizations
up to $K=34$, and
extrapolating these results  to infinite $K$,
we find the upper bounds $M^2\approx 10.75 \,g^2 N / \pi$
for $x=0$ and $M^2\approx 25.90 \,g^2 N / \pi$
for $x=1$. These are extremely close to the extrapolations from
Figs.6(a) and 6(b). This shows that, by judiciously truncating the space
of states, certain eigenvalues can be determined to a good accuracy with
relatively small diagonalizations.

Similar approximations work even better for the fermionic ground state
because it is purer than the bosonic ground state: for $x=0$ and $K=25$
this state has probability 0.999932 to consist of 3 partons, and probability
0.999997 to consist of 3 or 5 partons. Extrapolating these probabilities
to infinite $K$, we find 0.99983 and 0.999993, respectively.
With the 3-parton truncation and $K$ up to 75, we find that the
fermionic ground state has the extrapolated eigenvalue
$M^2\approx 5.72 \,g^2 N / \pi$ for $x=0$, and
$M^2\approx 26.05 \,g^2 N / \pi$ for $x=1$.
These values provide good upper bounds on the extrapolations from
Figs. 6(a) and 6(b).
Furthermore, the truncation that involves 3- and 5-parton states
is almost exact for the lowest fermionic eigenvalue, for all accessible
values of $K$. The advantage of this truncation is that we
can access higher value of $K$ (up to 49) than in the full diagonalization
and extrapolate more reliably.
We find that the lowest fermionic eigenvalue extrapolated
in this fashion is
$M^2\approx 5.70 \,g^2 N / \pi$ for $x=0$, and
$M^2\approx 25.94 \,g^2 N / \pi$ for $x=1$.
Good agreement of these values with Figs.~6(a) and 6(b) gives us some
confidence that our methods are consistent.

\bigskip
\centerline{\caps 5. Discussion}
\bigskip

One interesting property of our model is that, even in the limit
$m\to 0$, there are no massless bound states. This result was found
numerically in Ref.~[\DK] and can easily be explained analytically [\I].
For $m=0$ all the bound state masses are measured in units of $g$.
If we are interested only in the massless states, we can send
$g\to\infty$ so that the relevant action is
$$ S = \int d^2 x \Tr \left[
i\Psi^T\gamma^{0} \gamma^{\a} \partial_{\a} \Psi
+A_\alpha J^\alpha\right]\,. \eqn\strong$$
The left-moving currents $J^+_{ij}$ and the
right-moving currents $J^-_{ij}$ generate two independent level-$N$
Kac-Moody algebras. The gauge fields act as Lagrange multipliers
that enforce the zero-current conditions. Calculation of the central charge
of the Virasoro algebra
in such theories is well-known (see, for instance, Ref.~[\Polyakov]).
For $SU(N)$ the result is
$$ c=c_m - (N^2-1){K_m\over K_m+N}
,$$
where $c_m$ is the central charge before gauging, and $K_m$
is the level of the current algebra. In the theory with
an adjoint Majorana fermion $c_m=(N^2-1)/2$, and
we find $c=0$. This establishes
the absence of massless bound states. Similarly, in a gauge theory
coupled to a fundamental Dirac fermion, we have $c_m=N$ and $K_m=1$,
so that $c=1$. This proves the existence of one massless meson.
In the large-$N$ limit this phenomenon can be attributed to the breaking of
the $U(1)$ chiral symmetry [\Zh]. For comparison, note that
the $m=0$ version of the theory
\faction\ has no chiral symmetry because we are considering Majorana
fermions. This provides another physical reason for the absence of
massless bound states.

One of the motivations for coupling 1+1 dimensional QCD to the adjoint
matter is that the parton pair creation is not suppressed by a power
of $N$. This situation resembles large-$N$ QCD in higher dimensions,
where the quark creation is suppressed, but the gluon creation is not.
We find, however, that in the low-lying states the pair creation is
suppressed for dynamical reasons. Roughly speaking, in the low-lying
states the color flux tubes are not highly stretched, and it is
not energetically favorable for a flux tube to divide into 3 flux tubes
by creating a pair of partons in the middle. Indeed, even as $m$
is taken to zero, it costs some energy to create a pair of quanta
together with the associated flux tubes. This is why the first bosonic
excited state, which contains 4 partons, is considerably heavier than
the bosonic ground state, which contain 2 partons. However, as we pump
vibrational energy into a bound state, the
pair creation should become more favored.
This is confirmed by our computations.
It is remarkable, though, that the lowest states are so pure
that completely ignoring pair creation is an excellent approximation for them.
It is tempting to speculate that this is somehow connected with
the success of the ``valence approximation'' in 3+1 dimensional QCD
[\Wein].

Another important point concerns the rich structure of the excited states
found in 1+1-dimensional QCD with adjoint matter. This may be the
simplest class of models to exhibit a spectrum of Hagedorn type, with
an exponentially growing density of states. Since we have taken $N$
to infinity, all these glueball-like states are stable and can be
found as solutions of the linear equation \LS. For comparison,
the `t Hooft model has only one state per unit mass-squared.
These features are related to the presence of a deconfining phase
transition in the adjoint matter model, and its absence in the
`t Hooft model [\dkut].

Recently 1+1-dimensional large-$N$ QCD was connected with closed string
theory in a very precise fashion [\DG]. This was accomplished for
the pure glue theory, which is almost topological.
It is an interesting question, whether a continuum closed string
description exists for the more complicated adjoint matter model,
with its rich structure of physical states. The appearance
of the exponentially growing density of levels
suggests that such a theory should have a hidden transverse dimension.
In our \lc\ description this dimension manifests itself in the
fluctuations of the longitudinal momenta and of the number of partons.

Clearly, a lot remains to be understood about the
1+1-dimensional large-$N$ QCD coupled to adjoint matter.
We hope that this class of models bears some physical similarity
with higher-dimensional gauge theories. It may also serve as
a good test of the \lcq\ methods that are promising to become
a useful tool for studying the non-perturbative structure of the strong
interactions.

\ack
We are grateful to C. Callan, S. Dalley, M. Douglas,
D. Gross, D. Kutasov, A. Polyakov, S. Shenker, L. Susskind and E. Witten for
helpful discussions.
We also thank  Thinking Machines Corporation for use of their
computers.
The work of G. B. was partly supported by a US DOE Grant
DE-FG02-90ER40542 and by the Ambrose Monell Foundation.
I. R. K. and K. D. are supported in part by
NSF Presidential Young Investigator Award PHY-9157482 and
James S. McDonnell Foundation grant No. 91-48. I. R. K. is also
supported by DOE grant DE-AC02-76WRO3072 and
an A. P. Sloan Foundation Research Fellowship.
I. R. K. is grateful to the Institute for Theoretical Physics
in Santa Barbara for hospitality while this work was in progress.
During his stay there his research was supported in part by
the National Science Foundation under Grant No. PHY89-04035.

\refout

\endpage

\noindent {{\caps Table I:} $K=25$, $x=0$; the 3.0 bin includes
all states whose masses are $1.5-3.0$; etc.}
\medskip
\tablewidth=6.5in
\begintable
$M$  \| 3.0 | 4.5 | 6.0 |7.5 | 9.0 | 10.5 | 12.0 | 13.5 | 15.0 \cr
number of states \| 1|1|9|37|104|362|897|1668|2040 \cr
number of pure states\| 1|1|0|0|0|0|0|0|0\cr
average length \| 3.00|5.00|5.70|6.63|7.31|8.27|9.38|
10.50|11.73
\endtable

\bigskip

\noindent {{\caps Table II:} $K=25$, $x=1$}
\medskip
\tablewidth=6.5in
\begintable
$M$  \| 6.0 |7.5 | 9.0 | 10.5 | 12.0 | 13.5 | 15.0|16.5|18.0|19.5 \cr
number of states \| 1|1|5|17|56|131|296|580|942|1230 \cr
number of pure states\| 1|1|3|1|0|0|0|0|0|0\cr
average length \| 3.00|3.06|3.63|3.99|5.05|5.84|6.87|
7.88|9.07|10.36
\endtable

\bigskip

\noindent {{\caps Table III:} $K=24$, $x=0$}
\medskip
\tablewidth=6.5in
\begintable
$M$  \|4.5|6.0|7.5|9.0|10.5|12.0|13.5|15.0\cr
number of states \| 1|9|35|108|315|767|1229|1257 \cr
number of pure states\| 1|1|0|0|0|0|0|0\cr
average length \|2.05|5.36|6.43|7.21|8.23|9.32|10.48|11.74
\endtable

\endpage

\noindent{{\caps Table IV} }
\medskip
\tablewidth=6.5in
\begintable
\hfil\|\multispan{2}\tstrut\hfil $x=0$ \hfil \|
       \multispan{2}\tstrut\hfil $x=1$ \hfil \crthick
$K$  \| $M^2$ (full) | $M^2$ (2+4+6--bit) \|
$M^2$ (full) | $M^2$ (2+4+6--bit) \crthick
12\| 9.9710| 9.9711\| 19.7985| 19.7985\cr
14\|10.1034|10.1036\| 20.4120| 20.4120\cr
16\|10.2004|10.2008\| 20.8972| 20.8972\cr
18\|10.2742|10.2747\| 21.2923| 21.2923\cr
20\|10.3320|10.3326\| 21.6214| 21.6214\cr
22\|10.3783|10.3791\| 21.9004| 21.9004\cr
24\|10.4162|10.4171\| 22.1406| 22.1406\cr
$\vdots$\|$\vdots$| $\vdots$ \| $\vdots$ | $\vdots$\cr
$\infty$ \| 10.7 | 10.75 \| 25.9 | 25.90 \endtable

\noindent{{\caps Table V} }
\medskip
\tablewidth=6.5in
\begintable
\hfil\|\multispan{3}\tstrut\hfil $x=0$ \hfil \|
       \multispan{3}\tstrut\hfil $x=1$ \hfil \crthick
$K$  \| $M^2$ (full) | $M^2$ (3--bit) | $M^2$ (3+5--bit) \|
$M^2$ (full) | $M^2$ (3--bit) | $M^2$ (3+5--bit) \crthick
15 \| 5.5111 | 5.5119 | 5.5112 \| 21.1658 |21.1722|21.1659 \cr
17 \| 5.5388 | 5.5399 | 5.5389 \| 21.5335 |21.5427|21.5337 \cr
19 \| 5.5602 | 5.5617 | 5.5603 \| 21.8397 |21.8517|21.8400 \cr
21 \| 5.5771 | 5.5790 | 5.5772 \| 22.0996 |22.1143|22.0999 \cr
23 \| 5.5908 | 5.5930 | 5.5910 \| 22.3234 |22.3408|22.3238 \cr
25 \| 5.6021 | 5.6046 | 5.6022 \| 22.5187 |22.5385|22.5189 \cr
$\vdots$\|$\vdots$|$\vdots$| $\vdots$ \|$\vdots$| $\vdots$ | $\vdots$\cr
$\infty$ \| 5.7| 5.72| 5.70\| 25.9| 26.05|25.94\endtable

\endpage

\centerline{FIGURE CAPTIONS}
\bigskip

\item
{\bf Fig.1.} A glueball-like bound state of 6 partons.
\item
{\bf Fig.2.} The spectrum of fermionic states for $K=25$:
(a) $x=0$, $M <14$; (b) $x=1$, $M < 20$; mass $M$  is measured in units of
$\sqrt{g^2 N / \pi}$ and plotted vs. the expectation value
of the parton number.
\item
{\bf Fig.3.} The spectrum of bosonic states  for $K=24$:
(a) $x=0$, $M < 14$; (b) $x=1$, $M < 20$.
\item
{\bf Fig.4.} Logarithm of the density of states and a linear fit
for $K=25$ and $x=0$.
\item
{\bf Fig.5.} Average number of partons as a function
of mass for $K=24$ and $x=0$.
\item
{\bf Fig.6.} Fermionic and bosonic ground states and their
extrapolation towards infinite $K$: (a) $x=0$, (b) $x=1$.
\endpage

\bye